\documentclass[12pt,a4paper]{article}
\usepackage{amsmath}
\usepackage[dvipsnames]{xcolor}

\usepackage[left=3cm, right=3cm, top=3cm, bottom=3cm]{geometry}

\usepackage{eurosym}
\usepackage{lscape}

\usepackage{booktabs}
\usepackage{tabularx}
\usepackage{makecell}
\usepackage[flushleft]{threeparttable}
\usepackage{float}
\usepackage{caption}
\usepackage{longtable}

\usepackage{graphicx}
\usepackage{subcaption}

\usepackage{caption}
\usepackage{xcolor}

\usepackage{verbatim}

\usepackage[utf8]{inputenc}

\usepackage{setspace}
\usepackage{natbib}
\usepackage{lscape}

\usepackage{array}
\usepackage{arydshln}

\usepackage{abstract}

\usepackage{dcolumn}
\newcolumntype{,}{D{.}{.}{2}}

\usepackage[bottom]{footmisc}
\usepackage[skip=4pt, indent=0pt]{parskip}

\begin{document}
\setlength\parindent{0pt}
\thispagestyle{empty}
\renewcommand*{\thefootnote}{\fnsymbol{footnote}}

\begin{center}
\section*{Digital Transformation in Switzerland: The Current State and Expectations\footnote{This research has been funded by the Swiss National Science Foundation as part of the National Research Program NRP77 “Digital Transformation” under grant No. 187462. The authors are solely responsible for the analysis and the interpretation thereof.}}
\end{center}
\subsubsection*{Johannes Lehmann\footnote{University of Basel.} and \centering{Michael Beckmann\footnote{University of Basel, Institute for Employment Research (IAB), and Institute of Labor Economics (IZA).} }}

\bigskip
\bigskip
\bigskip

\begin{abstract}
This paper provides a comprehensive, descriptive overview of the current state of digital transformation in the Swiss economy and delineates areas that businesses should keep an eye on. Key findings illustrate that even established technologies are not universally adopted, that companies tend to overestimate their technological status compared to their competitors, and that it is important to have the appropriate technological know-how when introducing new technologies. In addition, companies expect changes in their work processes and employment conditions in connection with the digital transformation. Specifically, work tasks are expected to become more complex, diverse and varied. Employees' knowledge acquisition will gain in importance, especially in the form of formal further training and self-learning. Employees will also be more autonomous in making decisions about their jobs and working hours. 
\end{abstract}

\newpage
\pagenumbering{arabic}
\setcounter{page}{1}

\renewcommand*{\thefootnote}{\arabic{footnote}}
\setcounter{footnote}{0}

\section{Introduction}
The past few decades have witnessed significant changes in the business environment, largely driven by the almost universal adoption of novel technologies across the economy. The initial proliferation of basic information and communication technologies such as computers and cell phones, which are now in virtually universal use, was initially followed by the emergence of organizational tools such as enterprise resource planning or document management systems. More recently, even more sophisticated technologies like the Internet of Things, artificial intelligence, or blockchain applications have gained attention in the media and increasingly as practical applications in enterprises. The constant influx of new technological solutions has created fresh business opportunities while rendering many existing processes obsolete \cite[]{mullerFortuneFavorsPrepared2018}. Consequently, these developments can impact the structure of the work environment, the creation and improvement of existing and new products, as well as employee productivity (e.g., \cite{nambisanDigitalInnovationManagement2017, aboelmagedKnowledgeSharingEnterprise2018, papagiannidisSmartOfficesProductivity2020, braganzaProductiveEmploymentDecent2021}). Moreover, not only has the production side of businesses changed, but also consumer behavior has evolved as customers are increasingly taking advantage of the opportunities offered by new technologies. Consequently, businesses now face increasingly more informed and demanding customers (e.g., \cite{warnerBuildingDynamicCapabilities2019, fernandez-roviraDigitalTransformationBusiness2021, shakinaRethinkingCorporateDigital2021}).

Taken together, these arguments imply that digitization and IT innovation, in general, have a significant impact on the performance of businesses, as supported by an extensive literature (e.g., \cite{leidnerExaminationAntecedentsConsequences2010, luProactiveReactiveIT2010, beheraPerformanceMeasurementDue2015, nguyenInformationTechnologyAdoption2015, ellerAntecedentsConsequencesChallenges2020}).  Successful implementation and integration of new technologies into work processes, coupled with an appropriate level of technological expertise, can lead to a competitive advantage (e.g., \cite{bhattTypesInformationTechnology2005, franklinInnovationApplicationDigital2013}). However, while the literature generally agrees that an effective implementation of new technologies is crucial for business success and creates new opportunities, the question of how to effectively implement and select the most suitable technologies remains challenging and context-dependent for individual businesses \cite[]{nwankpaExploringEffectDigital2020}. These issues also hold true for the Swiss economy. \cite{beckAnalyseDigitalisierungSchweizer2020} note that Swiss businesses have not yet fully harnessed the potential for productivity gains arising along with digital transformation.
 
There is literature investigating the state of digitization in Swiss establishments, most notably the studies by \cite{beckAnalyseDigitalisierungSchweizer2020}, \cite{woerterInnovationSchweizerPrivatwirtschaft2020}, and \cite{speschaInnovationUndDigitalisierung2023}.\footnote{All these studies are based on the innovation and digitization surveys collected by the KOF, the economic research center of the ETH Zürich.} Our research complements these studies and provides additional insights in several ways. First, we complement \cite{beckAnalyseDigitalisierungSchweizer2020} by providing a detailed overview over applied technologies in Swiss businesses, where our investigation can be seen as an update to their findings relying on data from the year 2016. The investigated technologies are also not congruent as we include a detailed breakdown of how employees are equipped with digital devices.\footnote{Both \cite{speschaInnovationUndDigitalisierung2023} and \cite{woerterInnovationSchweizerPrivatwirtschaft2020} conduct investigations about the prevalence of different technologies, where the difference between the considered types of technologies is especially large.} Second, we extend the view on digitization presented in \cite{speschaInnovationUndDigitalisierung2023} and \cite{woerterInnovationSchweizerPrivatwirtschaft2020}, who examine topics such as investments in ICT, the provision of technology induced further training measures, and recruitments requiring digital skills, but do not pay much attention to the consequences of digitization on firm strategies and policies. In contrast, we explicitly consider the consequences of digitization on various outcomes such as outsourcing decisions (firm strategy), employee autonomy (firm policy), or employee requirements and well-being (e.g., demands for employee knowledge, psychological strain).\footnote{\cite{beckAnalyseDigitalisierungSchweizer2020} also look at potential effects of digitization with a focus on competitiveness and employment effects.} Adding these areas to the literature refines the perspective of the academic literature on the digital transformation in Switzerland and allows the proposition of more fine-tuned practical recommendations for Swiss establishments.

The remainder of this paper is structured as follows: Chapter 2 provides an overview of the applied dataset and the pursued descriptive methodological approach. Chapter 3 offers insight into the various technologies adopted by Swiss establishments and sheds light on the technological investments. Chapter 4 explores the importance of employee development accompanying the process of digitization. Chapter 5 investigates the establishments' self-assessment of their technological state and reveals that Swiss firms tend to overestimate their technological state. Chapter 6 examines the expected effects of the digital transformation on various outcome variables, such as outsourcing decisions, changes in work tasks, the evolving demands for employee knowledge, employee autonomy, and the psychological strain experienced by employees. Finally, chapter 7 presents concrete recommendations for action for Swiss establishments and concludes the paper.

\section{Methodology and Data}
This empirical investigation relies on data from an establishment-level survey, the \emph{Swiss Employer Panel} (SEP). The survey was handed out to a stratified random sample of 10'000 establishments. The contacted sample is representative of the Swiss establishment population with ten or more employees.\footnote{The federal statistical office of Switzerland provided the samples. The sample does not cover the public administration, farming, and mining sector (i.e., establishments active in sections A, B, O, T, or U of the NOGA 2008 classification were omitted). The establishments were contacted twice by letter.} An establishment is defined as a spatially separated organizational unit that either operates independently or is part of a larger organization. The final dataset includes data from 582 establishments.\footnote{This corresponds to a response rate of close to 6\%. A circumstance which can be attributed, at least partially, to the presence of Covid-19 during the survey periods.} It is worth noting that the number of responses may vary slightly in the subsequent descriptive analysis depending on the specific question. The displayed frequencies are based on the raw data without any imputed values.

To test the validity of the results, the data was weighted individually by employee size, industry, and region, as well as a combination of all three characteristics. Notably, the weighting process did not significantly alter the presented results. This suggests that despite the relatively low response rate, the data remains representative of the target population and lead to the decision to display the frequencies following from the raw data.

As this analysis focuses on the digital transformation in Switzerland, we look at the application of different technologies across establishments. The survey considers sixteen different technologies as presented in table \ref{tab:Tech}. In our analysis, we compare establishments by the number of technologies utilized, thereby employing the median as the threshold value. Establishments utilizing seven or more technologies are considered to have many technologies, while those employing six or less are classified as having few technologies. In terms of firm size classes, we follow the most commonly used classification in Switzerland and define establishments with 10 to 49 employees as small, those with between 50 and 249 employees as medium and establishments with 250 or more employees as large.\footnote{As mentioned earlier, micro-establishments with 9 or less employees are not present in the data set.}

To differentiate between sectors, establishments provided us with their NOGA classification. Establishments with a NOGA classification ranging between 1 and 43 are classified as belonging to the manufacturing sector, while establishments with a NOGA classification ranging between 45 and 99 are classified as belonging to the service sector. 

\begin{landscape}
\begin{table}[!ht] \centering
\caption{\textbf{Overview of the surveyed technologies}}
\label{tab:Tech}
    \begin{tabularx}{\linewidth}{XcX}\toprule
        \textbf{Technology} & \textbf{Abbreviation} & \textbf{Examples} \\ \hline
        Nonstationary IT-equipment & ~ & e.g., smartphones, tablets, notebooks \\ \hdashline
        Stationary IT-equipment & ~ & e.g., PCs, electronic cash registers, CAD systems \\ \hdashline
        Groupware/Collaborative applications & ~ & e.g., Slack, MS Teams, Zoom, Webex \\ \hdashline
        Cloud storage/computing & ~ & e.g., AWS, MS Azure, IBM Cloud, Google Cloud \\  \hdashline
        Enterprise Resource Planning & ERP & e.g., SAP, Net Suite, MS Dynamics \\  \hdashline        Customer Relationship Management & CRM & e.g., Salesforce, Oracle \\   \hdashline       
        Document Management Systems & DMS &  e.g., MS Sharepoint, Rubex \\  \hdashline
        Software or algorithms for IT-based process optimization & ~ & e.g., artificial intelligence, big data analytics \\  \hdashline
        Networking and control of machines and plants via the internet (cyber-physical systems) & CPS & ~\\  \hdashline
        Management Information System & MIS & e.g., Clarity Professional MIS \\  \hdashline
        Virtual boardroom & ~ & e.g., Sherpany, Dilligent Boards, iDeals, BoardEffect \\  \hdashline
        Internet of Things & IoT & e.g., IoT-sensors, RFID chips, e-grains, NFC \\  \hdashline
        Additive manufacturing processes & ~ & e.g., 3D printing \\    \hdashline      
        Robotics, automated transport or\\ production systems & ~ & e.g., industrial, service or mobile robots, drones \\  \hdashline
        Virtual/augmented reality & ~ & e.g., MS HoloLens \\  \hdashline
        Blockchain (distributed ledger technology)& ~ & e.g., cryptocurrencies, non-fungible tokens, smart contracts \\ \bottomrule
    \end{tabularx}
\end{table}
\end{landscape}

\section{State of digitization}
\subsection{Application of digital technologies}
\begin{figure}[ht!]
\caption{Usage of digital technologies}
\includegraphics[width=12cm, height=11.2cm]{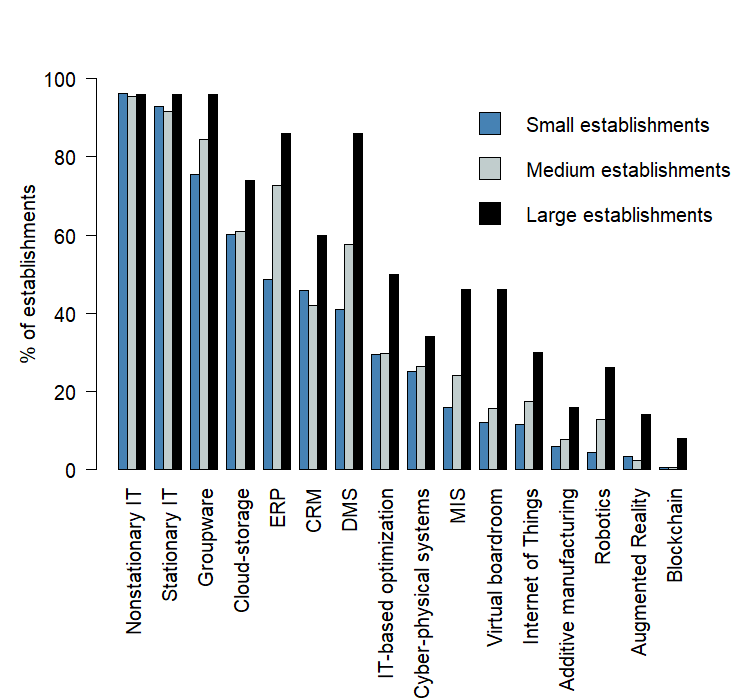}
\label{Tech_size}
\centering
\captionsetup{justification=centering,margin=1.5cm}
\caption*{\emph{Source: Swiss Employer Panel, 2022; own calculation and depiction.\\ Number of included observations: 412.}}
\end{figure}

A key indicator for assessing the level of digitization in specific establishments is the extent and intensity of their usage of digital technologies. By investigating the current utilization of digital technologies, we can gain insights into the overall digitization level within the Swiss economy \cite[]{beckAnalyseDigitalisierungSchweizer2020}. Figure \ref{Tech_size} illustrates the prevalence of the sixteen different hardware and software solutions which are present in the underlying survey and introduced in table \ref{tab:Tech} across the three different size classes of Swiss establishments.

Figure 1 shows that nearly all establishments possess some form of stationary and non-stationary electronic devices. Additionally, more than 80\% of establishments employ Groupware to facilitate communication among employees. Furthermore, approximately two-thirds of establishments utilize enterprise resource planning (ERP) software and cloud-storage solutions. Document management systems (DMS) and customer relationship management (CRM) tools are also widely adopted, with around half of establishments implementing them. The findings further indicate that one-third of establishments leverage IT-based optimization to enhance their business activities. Moreover, approximately one-fourth of establishments control machines or entire production lines through the Internet, known as cyber-physical systems, and employ management information systems (MIS). However, fewer than 20\% of establishments have embraced technologies such as virtual boardrooms or applications connected to the Internet of Things (IoT). Technologies such as autonomous functioning robotics, automated production lines or transport systems, in the figure subsumed as "Robotics", remain relatively uncommon, present in only one out of every ten establishments. The last three technologies queried in our study, i.e., additive manufacturing, virtual or augmented reality, and distributed ledger or blockchain technology, can still be considered as niche phenomena, utilized by 8\%, 4\%, and 1.5\% of establishments, respectively.

These findings demonstrate that while several digital technologies are relatively widespread, no single technology, apart from basic stationary and non-stationary IT, is universally adopted across all establishments. Furthermore, figure \ref{Tech_size} reveals a disparity in the adoption of technologies between establishments of different sizes. The clear result here is that all technologies are most often applied by large establishments. The only technology that small establishments use more often than medium establishments is CRM. The absolute as well as relative difference between the prevalence of the different technologies in large and small establishments are in some instances astonishing. In absolute terms the difference in the application of DMS (+ 45\%), ERP (+ 37.4\%) and virtual boardrooms (+ 34\%) are largest, with all technologies being more commonly employed in large establishments.

This pattern is also noted by existing literature. For instance, \cite{ellerAntecedentsConsequencesChallenges2020} find that small and medium-sized enterprises tend to be slower in digitization compared to large establishments. This discrepancy could be attributed to smaller establishments being more likely to have limited investment capacity and less experience, confidence, and competence with regard to implementing digital innovations (e.g., \cite{giotopoulosWhatDrivesICT2017, gruberProposalsDigitalIndustrial2019}; \cite{rafaelIndustryMaturityModel2020, lesoContributionOrganizationalCulture2023}). \cite{beckAnalyseDigitalisierungSchweizer2020}  identify a similar pattern in technology adoption across establishment size classes in their analysis of the Swiss economy in 2016. They note that there is a remarkable difference in the degree of digitization of larger establishments in comparison to smaller establishments and suggest that smaller establishments so far have have encountered challenges with regard to the digital transformation.

When comparing the usage of technologies across sectors, the differences are surprisingly small. The exceptions are technologies tied to the production of goods, such as robotics and additive manufacturing, which are more prevalent in the manufacturing sector.

\begin{figure}[ht!]
\caption{Number of implemented technologies}
\includegraphics[width=12cm, height=7.5cm]{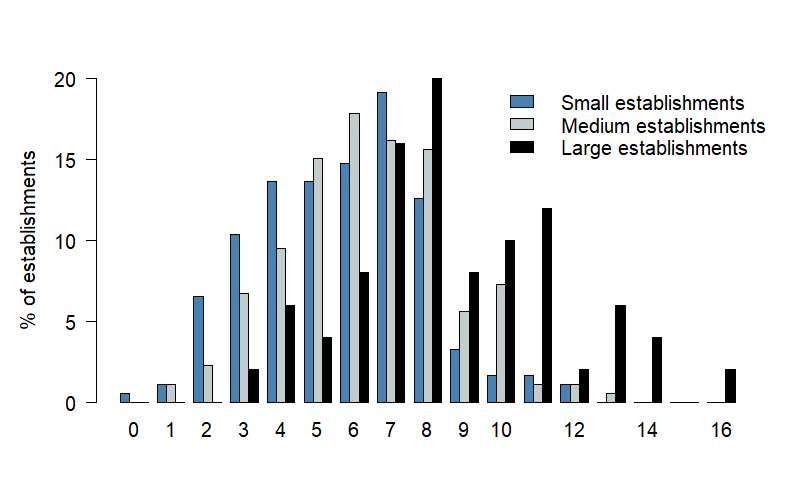}
\label{Tech_Anz_size}
\centering
\captionsetup{justification=centering,margin=1.5cm}
\caption*{\emph{Source: Swiss Employer Panel, 2022; own calculation and depiction.\\ Number of included observations: 412.}}
\end{figure}

Overall, the results identify basic technologies, which are adopted by nearly all establishments, namely stationary and non-stationary electronic devices. Additionally, there is a broad range of technologies that are widely used but not universally adopted, including ERP-systems, cloud-storage solutions, DMS, or CRM. A successful application of these technologies could serve as a distinguishing factor for establishments which are not at the frontier of the technological development and lead to a competitive advantage also in more conservative business areas. Finally, various technologies are identified as being applied exclusively by specialized establishments at the forefront of technological advancements.

Figure \ref{Tech_Anz_size} quantifies the number of different technologies applied in an establishment. The highest density is observed between five and eight of the queried technologies, encompassing over 60\% of the establishments. Nearly 85\% of the establishments utilize eight or fewer technologies. This indicates that most establishments complement their basic electronic devices with a couple of additional hardware or software solutions.

Consistent with the findings discussed in relation to figure \ref{Tech_size}, figure \ref{Tech_Anz_size} further confirms that large establishments tend to employ a greater number of technologies. Nevertheless, there are small (medium) establishments that utilize up to twelve (13) technologies in question. A classification of the establishments by sector reveals that, on average, manufacturing firms employ a greater number of technologies.

\subsection{Equipment with mobile devices}
\begin{figure}[ht!]
\caption{Share of employees with mobile devices}
\includegraphics[width=12cm, height=7.52cm]{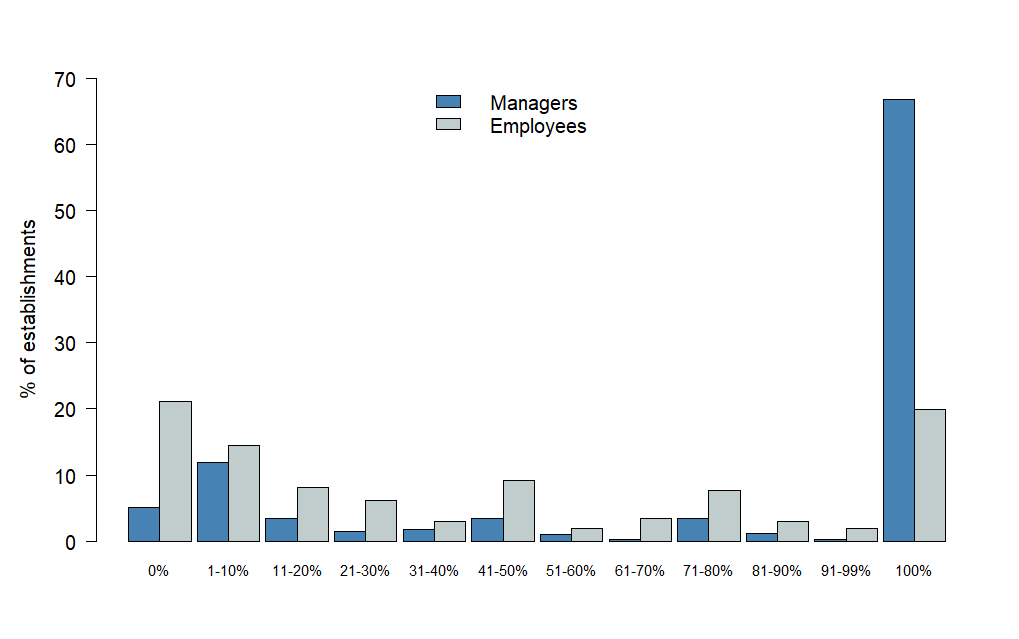}
\label{Mob_Proz}
\centering
\captionsetup{justification=centering,margin=1.5cm}
\caption*{\emph{Source: Swiss Employer Panel, 2022; own calculation and depiction.\\ Number of included observations: 410 (managerial employees); 406 (employees without managerial responsibilities).}}
\end{figure}

A further step of digitizing a business environment consists of equipping employees with mobile devices. This can enhance communication, facilitate information exchange within establishments, and, through this channel, increase business performance (e.g., \cite{stieglitzIncreasingOrganizationalPerformance2012, jeongEmployeesUseMobile2016}). Figure \ref{Mob_Proz} presents the proportion of managerial and non-managerial employees equipped with mobile devices. On average, 76\% of the managerial staff and 44\% of the non-managerial employees are provided with mobile devices. There is a considerable degree of heterogeneity among establishments, particularly regarding the provision of mobile devices to non-managerial employees. While over two-thirds of the establishments provide mobile devices to their entire management staff, 5\% of the establishments do not provide any mobile device to their management staff at all. The remaining 30\% of the establishments supply mobile devices to only a portion of their managerial staff, with the majority providing them to a minority of the management team.

Among establishments, approximately 20\% provide mobile devices to all non-managerial employees. A little over 20\% of the establishments do not supply any mobile devices to employees without managerial responsibilities. In the majority of the remaining 60\%, fewer than half of the employees are equipped with mobile devices. A similar share of establishments of different sizes provide mobile devices to at least part of their workforce. Moreover, there is no substantial difference between the service and manufacturing sectors in terms of equipping their management staff with mobile devices. However, employees without managerial responsibilities in the manufacturing industry are more likely to have access to business mobile devices compared to their counterparts in the service sector.

\subsection{Investments in ICT}
Before any technology can be implemented in an establishment, organizations must spend money and invest in these digital technologies. The extent of these investments, are the basis of ICT innovation. The investment decision is influenced by various factors, including the existing ICT expertise available in an organization, the competitive pressure, and the organizational agility \cite[]{nwankpaExploringEffectDigital2020}. Investing in digital technologies is crucial for organizations to maintain or create a competitive edge and meet evolving customer expectations regarding digital services and innovative products (e.g., \cite{davisUsingDigitalService2015, vonleipzigInitialisingCustomerorientatedDigital2017}). However, while the importance of digital investments for seizing innovation opportunities is evident, realizing sufficient benefits to offset the associated costs can be challenging. Successful implementation of new digital technologies often requires significant organizational changes, such as adapting the organizational culture, strategic decision-making processes, staffing, and communication practices \cite[]{gastaldiAcademicsOrchestratorsContinuous2015}.
Failures in the implementation and usage of digital technologies are more likely if establishments lack the necessary tools for proper implementation \cite[]{lokugeOrganizationalReadinessDigital2019}. In Switzerland, barriers to innovation investments include high costs, long amortization horizons, and uncertainty regarding the commercial success \cite[]{woerterInnovationSchweizerPrivatwirtschaft2020}. \cite{beckAnalyseDigitalisierungSchweizer2020} conclude that in Switzerland mostly establishments which are able to convert digital investments into new innovative services and products are the ones benefiting from investments into digital technologies, as they are able to cover the costs associated with the investments through the profits created by these new innovative services and products. In contrast, establishments focusing on cost reductions seem to benefit less.

\begin{figure}[ht!]
\caption{Share of investments in technological innovations}
\includegraphics[width=13cm, height=8.38cm]{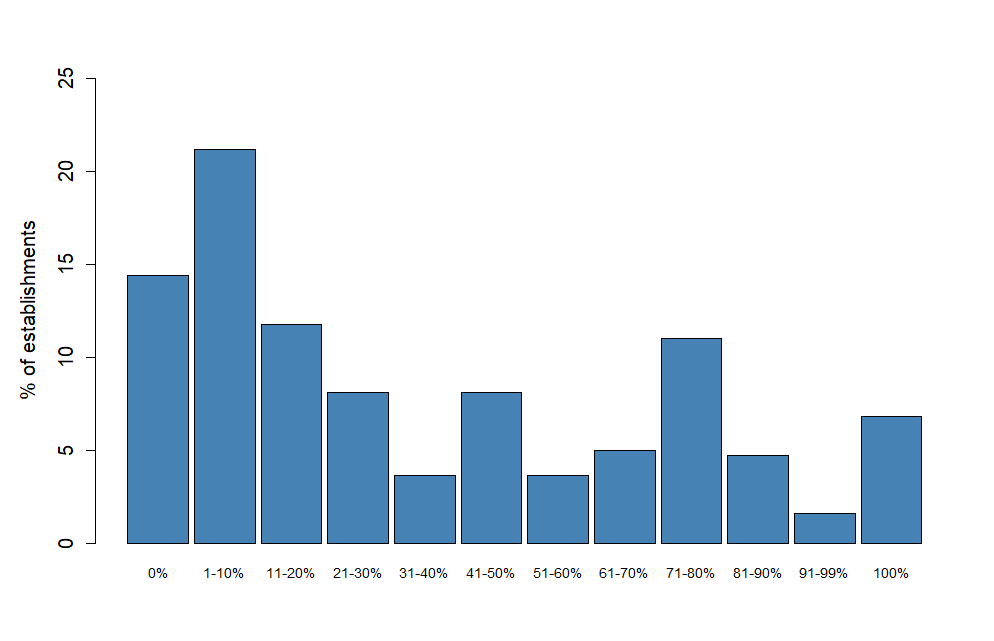}
\label{Inv_ICT}
\centering
\captionsetup{justification=centering,margin=1.5cm}
\caption*{\emph{Source: Swiss Employer Panel, 2022; own calculation and depiction.\\ Number of included observations: 382.}}
\end{figure}

Regarding investment patterns, the median establishment allocates approximately one-third of its total investments toward new technological devices and innovations. Figure \ref{Inv_ICT} demonstrates significant heterogeneity among establishments in this regard. Almost 15\% of the establishments did not incur any financial expenses related to investments in new technological devices or innovations in 2020. Approximately one-third of the establishments allocated between 1\% and 20\% of their total investment budget to this area. An additional 30\% of the establishments spent between 21\% and 70\% of their investment budget on technological devices and innovations. Almost 20\% of the establishments dedicated between 71\% and 99\%, while 7\% of the establishments exhausted their entire investment budget on technical devices and innovations.

Interestingly, despite significant differences between large and small firms in terms of technology adoption, the proportion of establishments investing in ICT is similar. This observation also reflects an overall lower investment volume of smaller establishments.\footnote{\cite{speschaInnovationUndDigitalisierung2023} also find no big difference in the share of ICT-investments between smaller and larger establishments.} Moreover, the distribution of investments into digital technologies is comparable between the manufacturing and the service sector, with the median investment volume being larger in the manufacturing sector.

\section{Employees development and digital technologies}
As already mentioned in the last chapter, the implementation of new technologies is not a straightforward process and does not guarantee success if the general conditions are not aligned. One crucial factor is the knowledge and expertise possessed by the establishment, or more specifically, its employees (e.g., \cite{acemogluChapter12Skills2011, dranoveTrillionDollarConundrum2014, woerterInnovationSchweizerPrivatwirtschaft2020}). In this sense, investing in new technologies is just an initial step. To effectively utilize the acquired technologies, establishments must have sufficient expertise related to these technological innovations. 
A substantial body of literature suggests that establishments need to create an appropriate environment and a corporate culture that fosters the development of employee knowledge if they want to reap the full benefit from the digital transformation (e.g., \cite{agostiniOrganizationalManagerialChallenges2019, assadiniaEffectsLearningOrientation2019, sousaSkillsDisruptiveDigital2019, agostiniDigitalizationInnovationProcess2020, gentile-ludeckeDoesOrganizationalStructure2020, zahoorTechnologicalInnovationEmployee2022}). One specific example is \cite{boothbyTechnologyAdoptionTraining2010}. The authors demonstrate that establishments that combine the implementation of digital technologies with strategic training gain more than establishments who only implement the digital technology. At least, they find that technological adopters in any case are more productive than establishments who do not make use of advanced technologies. Specifically, in the context of Switzerland, \cite{woerterInnovationSchweizerPrivatwirtschaft2020} propose that the lack of skilled professionals is a major hindrance to innovation.

In general, establishments have two options to equip their workforce with the necessary knowledge for the application of new technologies: hiring new employees with the required skills or providing training to the existing workforce to acquire the necessary new skills.

\begin{figure}[ht!]
\caption{Newly hired employees with IT-Skills}
\includegraphics[width=12cm, height=8.71cm]{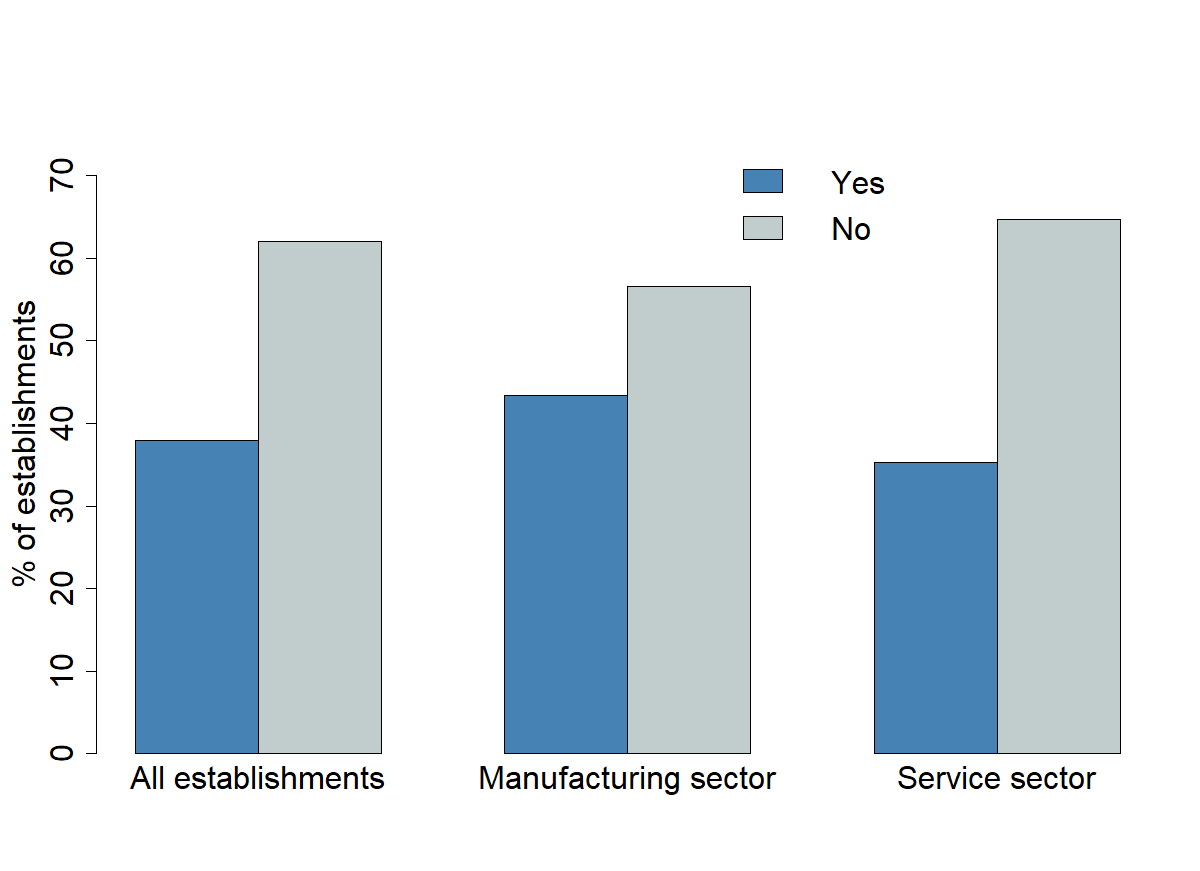}
\label{IT_Skills}
\centering
\captionsetup{justification=centering,margin=1.5cm}
\caption*{\emph{Source: Swiss Employer Panel, 2022; own calculation and depiction.\\ Number of included observations: 411.}}
\end{figure}

Figure \ref{IT_Skills} illustrates that 40\% of the establishments reported hiring employees with job profiles including skills such as programming, ICT development and data analytics in the past two years. On average, one in ten recruitments was associated with at least on of these characteristics. Furthermore, the figure shows that a higher proportion of establishments in the manufacturing sector hired at least one new employee with such knowledge.

Recruitments related to ICT skills are more prevalent in larger establishments, but this could also be influenced by the fact that larger establishments are generally more likely to hire new employees. Additionally, establishments that utilize an above-average number of technologies also reported a higher number of recruitments connected to ICT skills.

\begin{figure}[ht!]
\caption{Further training associated with digital development}
\includegraphics[width=10cm, height=8cm]{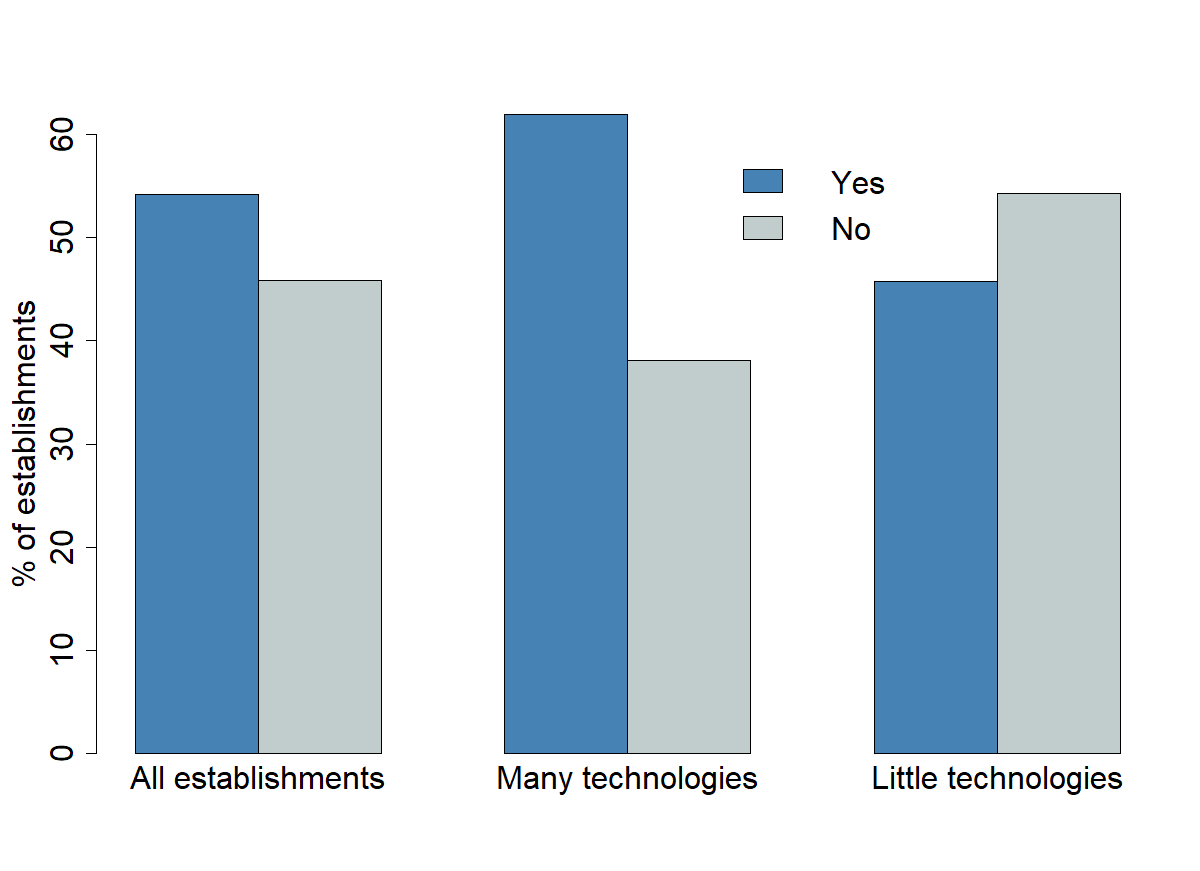}
\label{WB_Tech}
\centering
\captionsetup{justification=centering,margin=1.5cm}
\caption*{\emph{Source: Swiss Employer Panel, 2022; own calculation and depiction.\\ Number of included observations: 338.}}
\end{figure}

An alternative approach to hiring personnel with skills related to technological advancements is to provide training for existing employees. As shown in figure \ref{WB_Tech}, almost 55\% of the establishments reported investing in technology-induced training measures. On average, establishments that supported such training measures allocated 20\% of their total training budget to initiatives connected to technological change. Additionally, figure \ref{WB_Tech} shows that establishments that are more digitized, as measured by the number of technologies they employ, are more likely to financially support training measures associated with digital technologies. 62\% of the more digitized establishments financed technology-induced further training and only 46\% of the less digitized establishments did.

There is no significant difference in the probability of providing such training measures between establishments of different sizes, as well as between establishments operating in the service or manufacturing sector.

\section{Self-assessment of the technological state}
\begin{figure}[ht!]
\caption{Self-assessment of technological state}
\includegraphics[width=12cm, height=8.71cm]{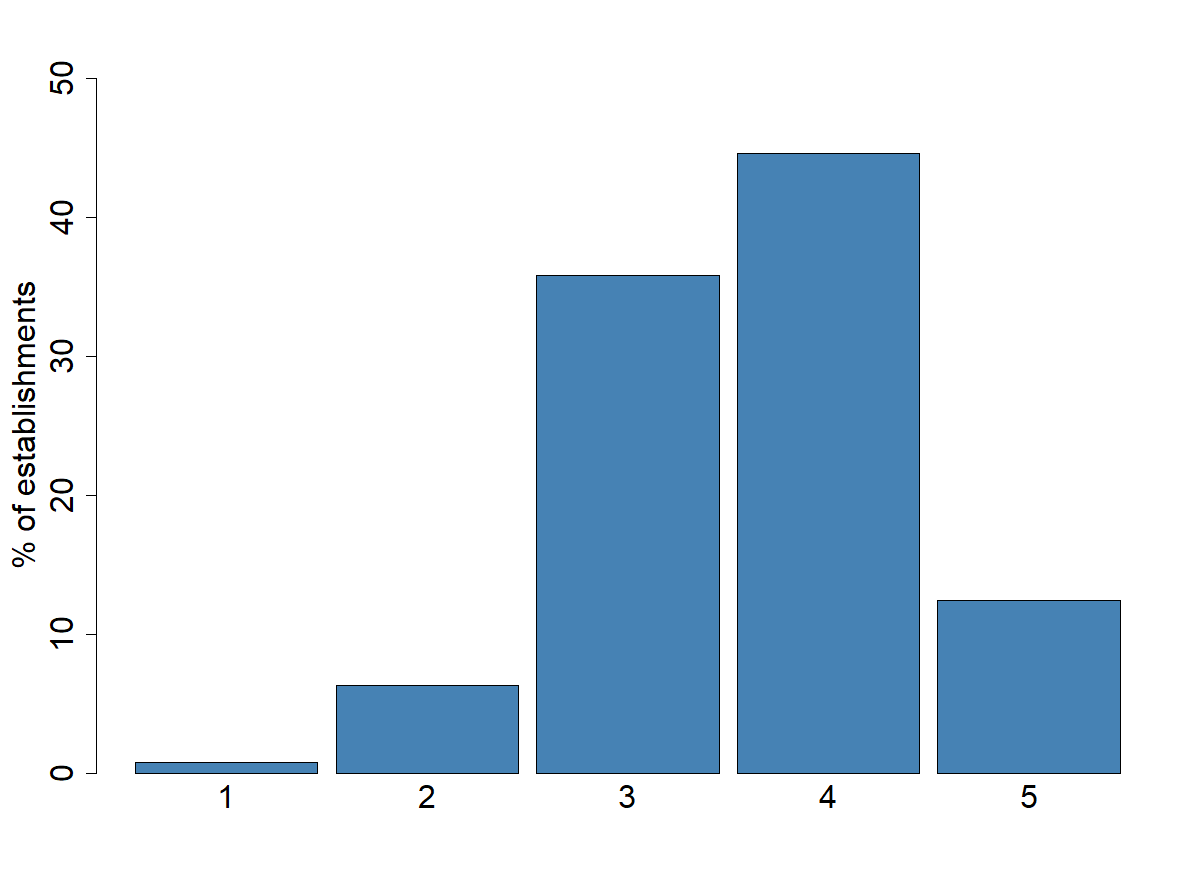}
\label{TechS}
\centering
\captionsetup{justification=centering,margin=1.5cm}
\caption*{\emph{Source: Swiss Employer Panel, 2022; own calculation and depiction.\\ Number of included observations: 410.}}
\end{figure}

Figure \ref{TechS} presents the self-assessment of establishments regarding the technological state of their equipment compared to their direct competitors. The ordinal scale ranges from completely out of date (1) to state-of-the-art equipment (5). A value of 1 or 2 indicates a below-average self-evaluation, 3 an average self-evaluation, and the values 4 and 5 indicate an above-average self-evaluation. The results show that almost 60\% of the establishments believe their technological state is above average, while less than 8\% assess their technological state as below average. Approximately one-third of the establishments consider their technological state to be average. Overall, this result shows that most establishments positively assess the overall state of their company's equipment.

This result can be interpreted in two ways. On the one hand, it can be seen as reassuring that the majority of establishments feel confident in their ability to handle the digital transformation and perceive their technological state as supportive of their business activities. On the other hand, the result suggests that Swiss establishments tend to overestimate their digital capabilities. It is mathematically impossible that the technological state is above average in 60\% of the establishments, indicating a potential overestimation. This overconfidence could lead to under-investment in new technologies, ultimately hampering the establishments' competitiveness in the market.

The finding that establishments tend to overestimate their position in the market is consistent with the results obtained in other papers. Several studies have shown that establishments have a tendency to overestimate the satisfaction of their customers \cite[]{suchanekCustomerSatisfactionDifferent2018} as well as their own performance and financial management \cite[]{knottsInternalVsExternal2012} in comparison with their competitors. Obviously, this overestimation bias does also refer to the assessment of technological capabilities.

\begin{figure}[ht!]
\caption{Self-assessment of technological state: By usage of technologies}
\includegraphics[width=12cm, height=8cm]{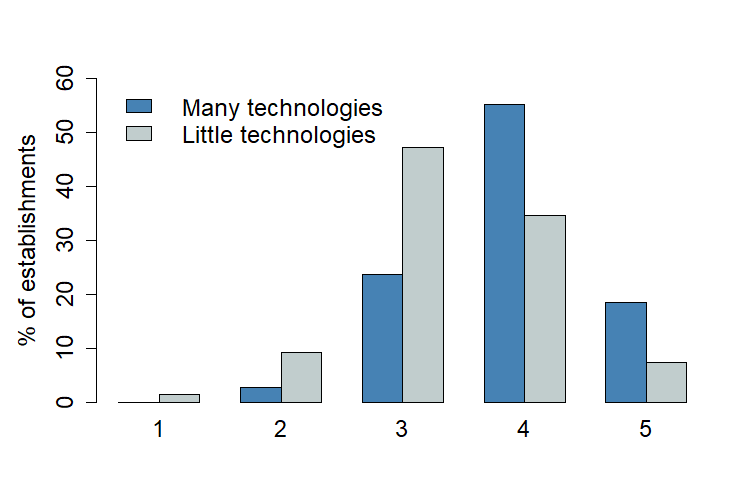}
\label{TechS_Anz}
\centering
\captionsetup{justification=centering,margin=1.5cm}
\caption*{\emph{Source: Swiss Employer Panel, 2022; own calculation and depiction.\\ Number of included observations: 410.}}
\end{figure}

Figure \ref{TechS_Anz} provides further insights into the self-assessment of establishments' technological state, similar to figure \ref{TechS}, but divided into two groups based on the number of technologies they use. Among establishments that utilize a high number of technologies, nearly 75\% perceive their technological state as above average, while less than 3\% consider it below average. Even among establishments which apply a comparatively low number of technologies, over 40\% still describe their technological state as above average, with only a little more than 10\% assessing it as below average. This suggests that the level of technology adoption and usage has a significant impact on establishments' self-perception of their technological capabilities.

A comparison of establishments by size reveals no strong divergence. Small establishments are even the ones assessing their technological state the highest. On average, large establishments rate their technological state with an average score of 3.6, medium establishments rate it at 3.5 and small establishments at 3.7. This suggests that the self-assessment of technological state, as shown in figure \ref{TechS_Anz}, is not primarily influenced by the size of the establishment but really by the decision of how many technologies are being applied. 

Additionally, there is no significant divergence in the results when comparing establishments in the manufacturing sector (mean score = 3.60) to those in the service sector (mean score = 3.65). This indicates that the sector in which establishments operate does not have a substantial impact on their self-perceived technological state. 

\section{Expected effects of the digital transformation}
The ongoing digital transformation brings about extensive consequences for various aspects of establishments' business activities. Changes in workflows result in altered conditions and requirements for employees. This section explores changes establishments expect due to the digital transformation. It encompasses a wide range of areas, including potential effects on task outsourcing, modifications of work tasks, shifts in employee knowledge requirements, changes in employee autonomy, and the impact on employee mental workload.

\subsection{Effect on outsourcing}
The emergence of digital technologies has a profound impact on business decision-making, among others in determining whether establishments should internally handle certain work tasks or outsource them. From a theoretical economic perspective, the internal and external transaction costs determine this decision. Lower internal transaction costs (relative to external transaction costs) make it beneficial to carry out work tasks in-house, while lower external transaction costs (relative to internal transaction costs) make outsourcing more profitable.\footnote{The fundamentals of transaction cost theory was developed by \cite{coaseNaturFirm1937}.}

The digital transformation significantly influences both external and internal transaction costs (e.g., \cite{abramovskyOutsourcingOffshoringBusiness2006, chenImpactInformationCommunication2016}). Examples of decreases in external transaction costs are simplified communication with suppliers or an easier access to information via the Internet. Conversely, internal transaction costs can be reduced through improved communication within establishments using tools such as enterprise resource planning (ERP) systems and virtual boardrooms. Consequently, the ratio
of how external and internal transaction costs change in response to new technologies ultimately determines whether integration or outsourcing is favored. Understanding these dynamics is crucial for establishments seeking to optimize their operational efficiency amidst the ongoing digital transformation.

Research findings regarding the effects of digital transformation on integration and outsourcing decisions vary across different sectors, type of establishments and even specific work tasks. For instance, \cite{demirbasImpactDigitalTransformation2018} demonstrate that the digital transformation has led to increased outsourcing in the financial service sector. In contrast, \cite{felserDigitalizationEvolvingIT2020} reveal that the adoption of new technologies enabling flexible and cost-efficient production has prompted the backshoring of previously outsourced ICT services in the German automotive industry. \cite{abramovskyOutsourcingOffshoringBusiness2006} examine the whole private sector and find that ICT reduces external coordination costs more than internal coordination costs. Consequently, they find that technology usage is positively associated with outsourcing and offshoring. Finally, \cite{chenImpactInformationCommunication2016} reveal a complementary relationship between intra-firm trade and outsourcing for multinational enterprises, as ICT reduces the communication costs with external trading partners as well as with foreign subsidiaries. 

\begin{figure}[ht!]
\caption{Effect on proportion of contracts awarded to external contractors }
\includegraphics[width=10cm, height=9.71cm]{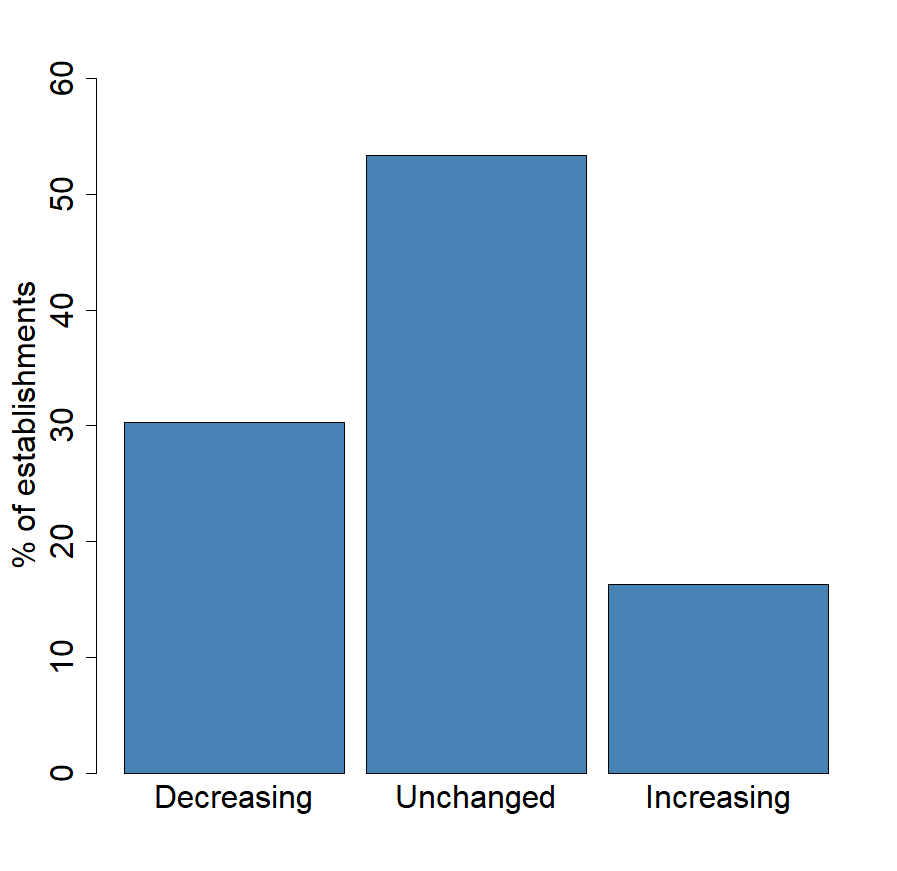}
\label{Out}
\centering
\captionsetup{justification=centering,margin=1.5cm}
\caption*{\emph{Source: Swiss Employer Panel, 2022; own calculation and depiction.\\ Number of included observations: 399.}}
\end{figure}

Figure \ref{Out} illustrates that more than 50\% of the surveyed establishments anticipate no change in the proportion of contracts awarded to external contractors. Only 16\% of establishments expect an increase, while nearly one third of the establishments anticipate a decrease. This suggests that a minority of establishments (16\%) anticipate a greater decrease in external transaction costs compared to internal transaction costs, while over 30\% expect a greater decrease in internal transaction costs relative to external transaction costs.

\begin{figure}[ht!]
\caption{Effect on proportion of contracts awarded to external contractors: Sorted by self-assessment of technological state}
\includegraphics[width=12cm, height=9.926cm]{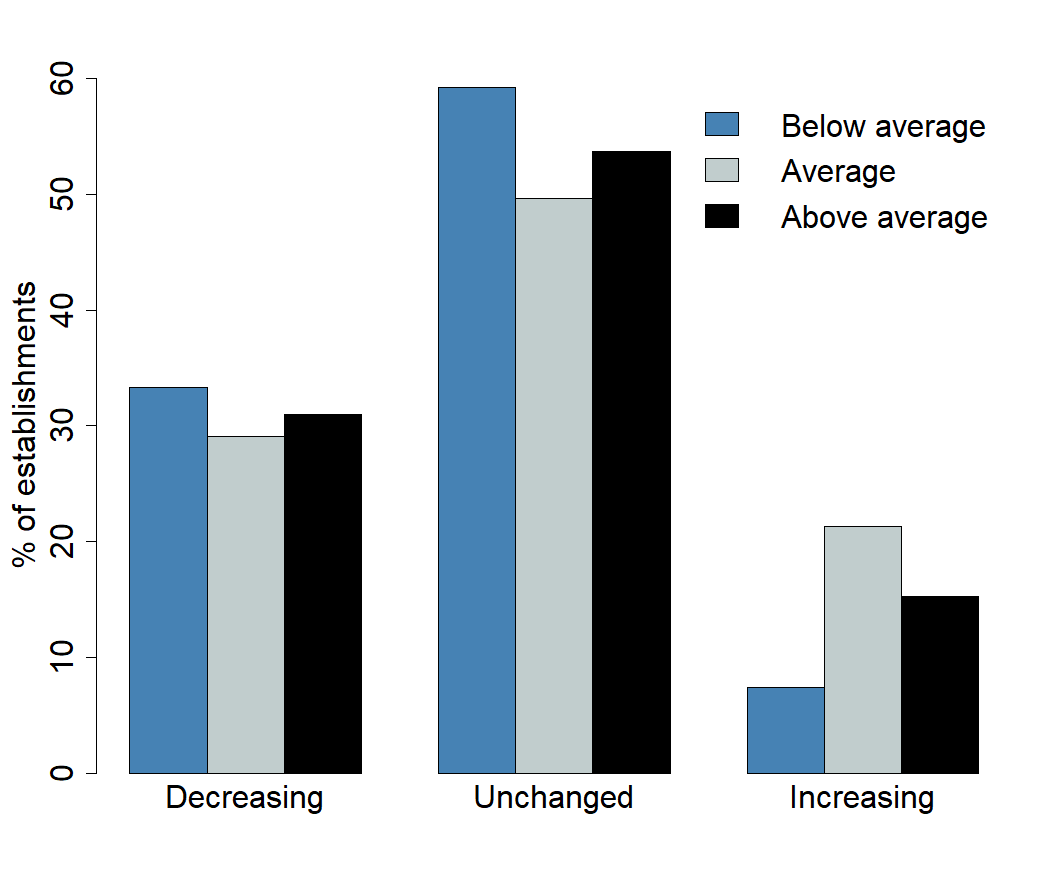}
\label{Out_TS}
\centering
\captionsetup{justification=centering,margin=1.5cm}
\caption*{\emph{Source: Swiss Employer Panel, 2022; own calculation and depiction.\\ Number of included observations: 399.}}
\end{figure}

Figure \ref{Out_TS} further examines the expected effect of digital transformation on the proportion of contracts awarded to external contractors, this time sorted by the self-assessment of technological state. It is observed that establishments with an average technological state are most likely to anticipate an increase in outsourcing. On the other hand, establishments with a below-average technological state are more inclined to expect no change and are very unlikely to expect an increase in outsourcing. This finding is interesting as it suggests that these establishments, do not appear to anticipate a growing need for external knowledge, even though, in their own assessment, they do not fully exploit the potential benefits of the digital transformation. Establishments that assess their technological state as above average fall in the middle across all three categories. Overall, therefore, the decreasing effect on the proportion of contracts awarded to external contractors appear to dominate the corresponding increasing effect, suggesting that Swiss firms a more likely to grow than to decline. This may be indicate that current technological progress is different from the consequences of technological change for the boundaries of a firm in the time between 1990 and the early years of the new millennium, where the opposite could be observed, on average.

\subsection{Effect on work tasks}
Digitization has the potential to significantly impact the nature and variety of work tasks. We examine three potential shifts in the type of work tasks as a result of digitization. Firstly, we investigate how digitization affects the diversity of work tasks. On one hand, technological advancements that streamline the workflow may lead to a decrease in task variety, similar to the specialization and monotonous work observed during the second industrial revolution. On the other hand, if new software solutions take over routine tasks and employees can focus on exceptional cases or interact with a wider range of work areas, task variety could increase.

Secondly, we explore the effect of digitization on the number of work tasks requiring complex problem-solving. Similarly to the argumentation concerning the diversity of work tasks, the impact could go in both directions. Simplification of work tasks through digital devices and software might lead to a decline in the number of complex tasks. Conversely, digitization may introduce more complicated tasks, resulting in an increase in their number.

Lastly, we examine the expected effect of digitization on the number of work tasks to be completed simultaneously. The underlying question here is whether digitization accelerates workflow and increases the scope of work for employees. If the speed and scope of workflows increase, employees may need to respond to multiple tasks simultaneously, thus increasing the number of tasks they must handle concurrently.

While the effects of digitization on work tasks are not straightforward, the literature tends to suggest an overall increase in diversity, complexity, and the number of simultaneous tasks. According to \cite{agostiniDigitalizationInnovationProcess2020}, increased automation expands the scope of responsibility and, as a consequence, the diversity of work tasks as understanding the connection between different processes becomes more important. \cite{rintalaTechnologicalChangeJob2005} and \cite{johnsonReviewAgendaExamining2020} argue that work tasks and jobs, in general, are becoming more complex and require autonomous knowledge work. However, \cite{marlerInformationTechnologyChange2012} find that complexity increases primarily for lower-level employees, while management and technical service jobs experience a lesser increase. There are counterarguments as well, such as those in \cite{peetersBetterWorseImpact2022} who show a negative relationship between the use of robotic process automation and task variety.

\begin{figure}[ht!]
\caption{Effect on work tasks}
\includegraphics[width=15cm, height=7.945cm]{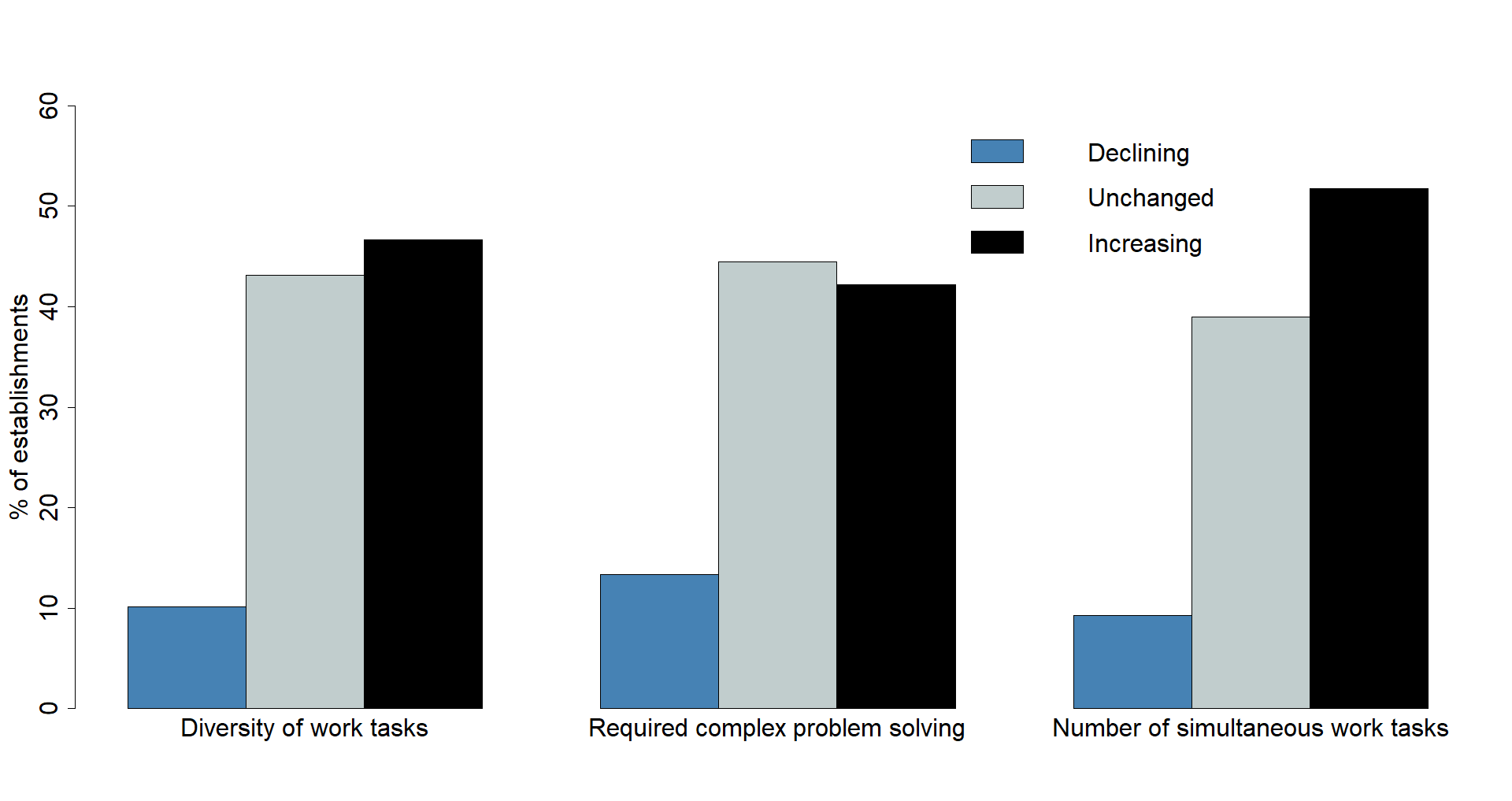}
\label{Work_Tasks}
\centering
\captionsetup{justification=centering,margin=1.5cm}
\caption*{\emph{Source: Swiss Employer Panel, 2022; own calculation and depiction.\\ Number of included observations: 403 (diversity of tasks); 398 (required complex problem solving); 400 (number of simultaneous tasks).}}
\end{figure}

Figure \ref{Work_Tasks} presents the effects establishments expect of digitization on the diversity of work tasks, the number of tasks requiring complex problem-solving, and the number of tasks to be carried out simultaneously. The general pattern is consistent across all three items. Firstly, only a small minority (around 10\%) of establishments expect a decrease in any of the three variables. Secondly, a significantly larger share (between 40\% and 50\%) of establishments expect an increase in each dimension. Lastly, approximately 40-45\% of establishments anticipate no change. The most notable difference is that more establishments expect no change for complex problem-solving tasks, while for the other two variables, more establishments expect an increase. However, the underlying trend is clear: very few establishments expect a simplification of work tasks or workflow, a substantial number expect no significant change, and many anticipate an increase in the complexity and pace of work tasks.

A further analysis shows that large establishments, in contrast to the baseline, are more likely to expect an increase in the diversity of work tasks and tasks requiring complex problem-solving. However, there is no significant difference in the expectations of small and large establishments regarding the number of tasks to be carried out simultaneously.

When examining establishments by sector, those in the service sector are more inclined to anticipate no change in all three variables. Interestingly, establishments in the manufacturing sector are more likely to expect both an increase and a decline in all three variables. It is worth noting that the number of manufacturing establishments expecting a decline are still a clear minority.

Furthermore, establishments that utilize a higher number of technologies are more likely to expect an increase in all three variables. Conversely, establishments with a lower number of technologies are more likely to anticipate a decrease in all three variables. Again, establishments expecting a decline remain a clear minority.

\subsection{Effect on requirements of employee knowledge}
This section delves into the effects of digitization on the requirements for employee knowledge. As previously discussed in section 4, the literature emphasizes the importance of appropriate knowledge within establishments for a successful implementation of new digital technologies. To comprehensively capture this dimension, we examine three distinct effects.

Firstly, and very generally, we are interested in understanding the expected effect of digitization on the demand for employee training. This encompasses both on-the-job training and additional training measures conducted in specialized institutions. Secondly, we investigate whether establishments perceive a shift in the responsibility for employees to keep their knowledge up to date. This would imply that employees take on a more proactive role in deciding the type of knowledge needed, rather than relying on the management for this decision. Lastly, we explore the impact of digitization on the importance of certifications, namely vocational training qualifications or academic degrees.

\begin{figure}[ht!]
\caption{Effect on requirements of employee knowledge}
\includegraphics[width=15cm, height=7.945cm]{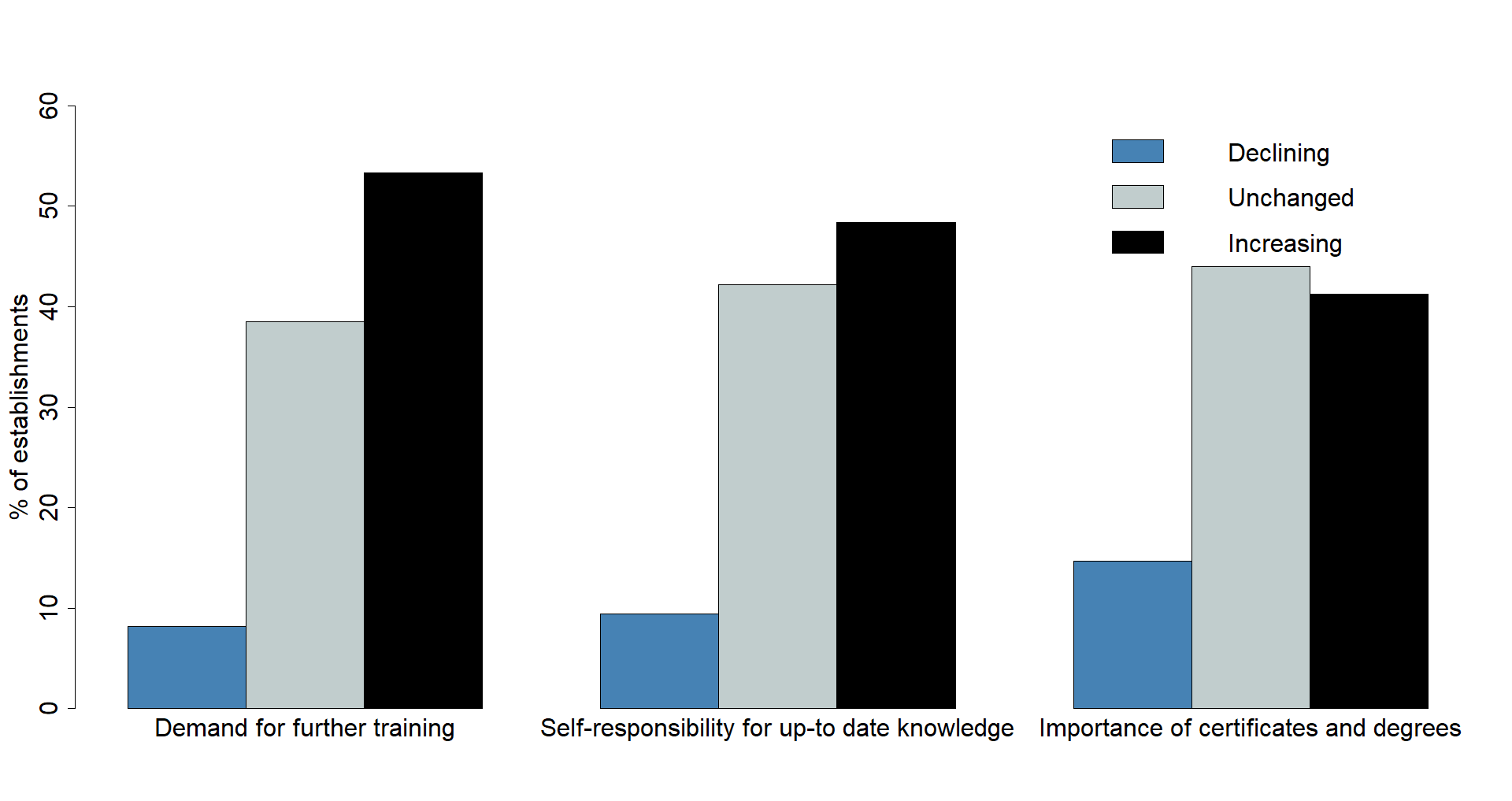}
\label{Know}
\centering
\captionsetup{justification=centering,margin=1.5cm}
\caption*{\emph{Source: Swiss Employer Panel, 2022; own calculation and depiction.\\ Number of included observations: 405 (further training); 403 (self-responsibility); 402 (certificates).}}
\end{figure}

Figure \ref{Know} presents the results pertaining to these three questions. Upon initial observation, it is apparent that while there is considerable heterogeneity among the three variables, a clear underlying trend emerges. Namely, only a small minority anticipates a decrease in the importance of any of the three variable. Approximately 40\% of the establishments do not expect any change in relation to each of the three variables. Almost 55\% of the surveyed establishments expect an increase in the overall demand for employee training, nearly 50\% expect an increase in the need for employees to independently keep their knowledge up to date, and slightly over 40\% anticipate an increase in the importance of vocational training certificates or academic degrees.

Notably, there is an approximate 15-percentage-point gap between the share of establishments expecting an increasing importance of further training measures and those expecting an increasing importance of certificates and degrees. This suggests that a sizeable portion of establishments does not believe that academic and vocational training adequately meet their specific knowledge needs. The fact that nearly 50\% of the establishments foresee a greater role for employees in selecting the type of knowledge they need could imply a shift in responsibilities. However, it may also indicate that managers recognize the highly specialized nature of the required knowledge, implying that employees are more aware of their own needs for successful navigation within the work environment than the management could ever be. Overall, figure \ref{Know} shows that most establishments agree with the notion of the literature and expect an increase in the demands for employee knowledge.

Turning our attention to establishment subgroups shows that large establishments are more likely to anticipate an increase in all three variables. The difference is particularly notable when it comes to the self-responsibility of employees to keep their knowledge up to date. Similarly, establishments that utilize a higher number of technologies are also more inclined to expect an increase in the importance of all three variables. The disparity is particularly pronounced in relation to the demand for further training (with a difference of +16 percentage points) and the self-responsibility of employees to keep their knowledge up to date (with a difference of +18 percentage points). When dividing the establishments across sectors, manufacturing firms exhibit a higher likelihood of expecting an increase in the importance of vocational training qualifications and academic degrees, with a difference of 18 percentage points compared to establishments in the service sector.

\subsection{Employee Autonomy}
Digitization may have a notable impact on employee autonomy, manifesting in various ways. One aspect that has received considerable attention is the effect on centralization and decentralization in organizational and job design, which is triggered by the introduction of new information and communication technologies \cite[]{bloomDistinctEffectsInformation2014a}. The adoption of communication technology is associated with increased centralization, concentrating decision-making power at higher levels of the organizational hierarchy. Conversely, the implementation of new information technologies tends to promote decentralization, granting employees greater decision-making authority. Additionally, digital technologies can enable employees to work from different locations, beyond the confines of the traditional office. Moreover, more independent work processes can allow employees the flexibility to determine the start and end times of their workday independently. Both trends - autonomy regarding working place and working time - have been found to be associated with an increase in job satisfaction and productivity (e.g., \cite{bloomDoesWorkingHome2015a, angeliciSmartworkingWorkFlexibility2020, bloomHowHybridWorking2022a}).

\begin{figure}[ht!]
\caption{Effect on autonomous choice of workplace and work time}
\includegraphics[width=10cm, height=10.26cm]{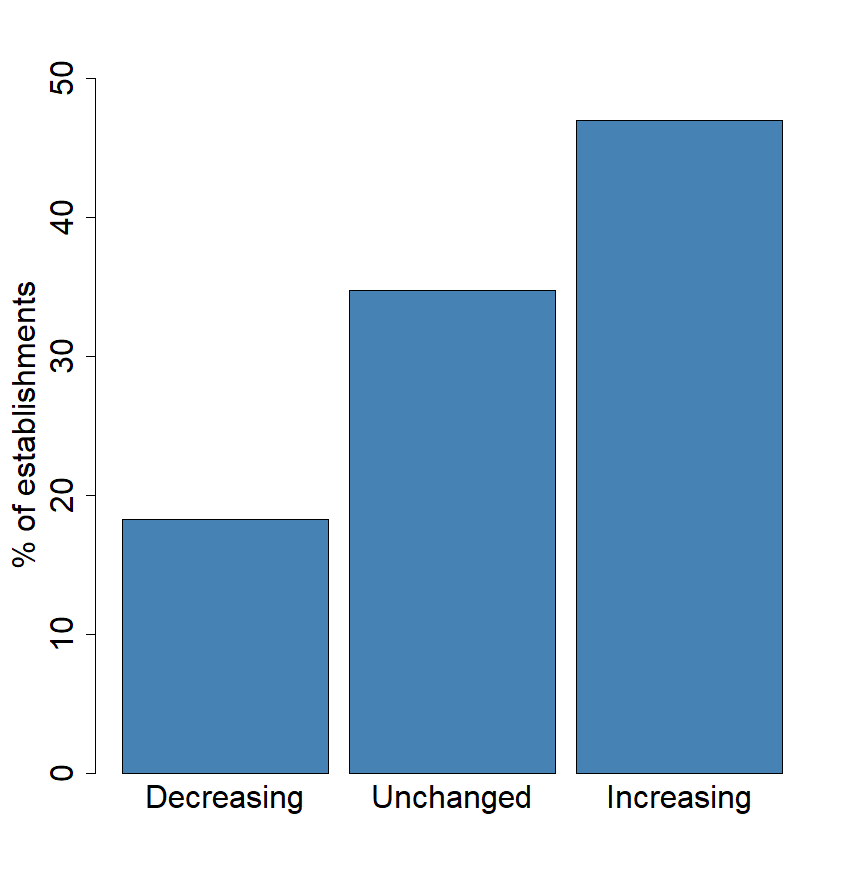}
\label{Aut}
\centering
\captionsetup{justification=centering,margin=1.5cm}
\caption*{\emph{Source: Swiss Employer Panel, 2022; own calculation and depiction.\\ Number of included observations: 400.}}
\end{figure}

Figure \ref{Aut} illustrates the anticipated effects of using digital technologies on the options available to employees regarding their own choice of working time and workplace. Approximately 20\% of the establishments expect a decrease in autonomy in this regard, while around 35\% anticipate no change. Notably, nearly half of the establishments expect an increase, signaling a growing recognition of the potential for enhanced employee autonomy in terms of choosing their working time and workplace.

The expectations of increased workplace and working time autonomy vary among different types of establishments. Large establishments, those operating in the service sector, and those applying a higher number of technologies are more likely to anticipate an increase in employee autonomy.

\begin{figure}[ht!]
\caption{Autonomous choice of workplace and work time: Sorted by self-assessment of technological state}
\includegraphics[width=10cm, height=10.26cm]{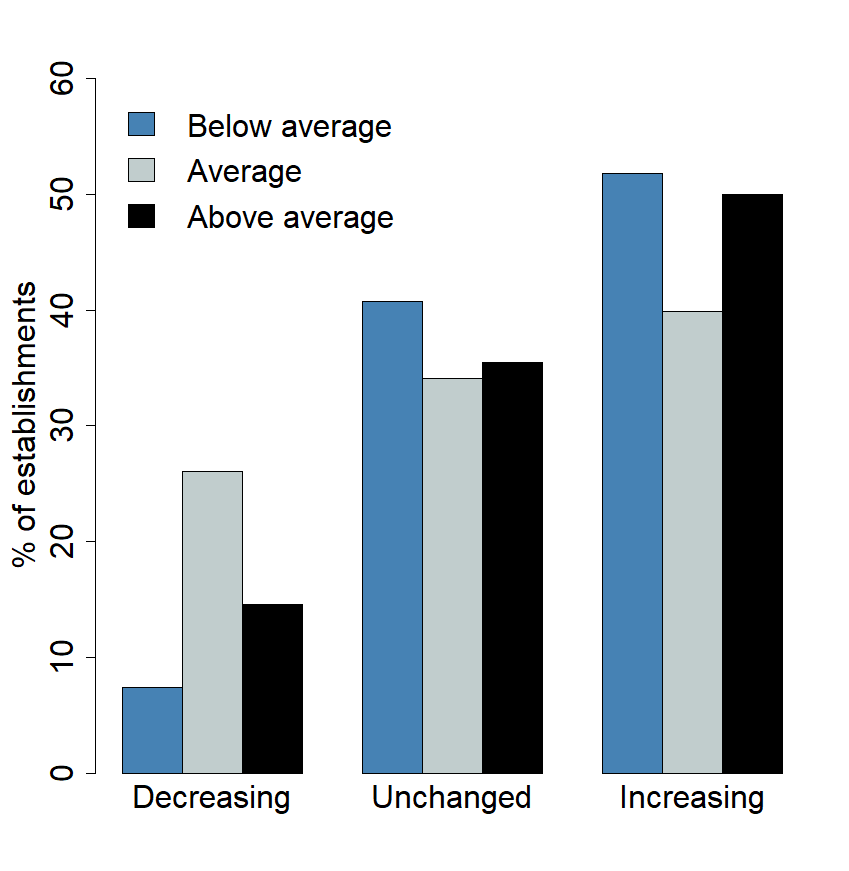}
\label{Aut_TS}
\centering
\captionsetup{justification=centering,margin=1.5cm}
\caption*{\emph{Source: Swiss Employer Panel, 2022; own calculation and depiction.\\ Number of included observations: 400.}}
\end{figure}

Figure \ref{Aut_TS} provides insights when the responses are categorized based on the self-assessment of the technological state. Establishments that assess their technological state as average are more likely to expect a decrease in autonomous choice, with 26\% of these establishments selecting this option. In contrast, only 40\% of these establishments expect an increase. Conversely, establishments that assess their technological state as below average demonstrate a different pattern. Only 7\% of these establishments expect a decrease, while 52\% anticipate an increase in the autonomous choice of workplace and work time. Establishments with an above-average self-assessment of their technological state fall in the middle of the other two categories.

\subsection{Effect on mental workload}
As discussed in the previous sections, many establishments anticipate changes in the type and flow of work tasks, along with an increased demand for employee knowledge through further training and self-learning. Additionally, the usage of technical devices, such as mobile phones and access to work-related information from home, can blur the boundaries between private life and work time, potentially affecting the mental strain on employees.

Studies by \cite{chesleyInformationCommunicationTechnology2014} and \cite{maierInformationTechnologyDaily2015} have linked the use of information technology in the workplace to higher levels of exhaustion and stress among employees. One underlying factor is the intensification of work processes, often accompanied by increased multitasking. \cite{barleyEmailSourceSymbol2011} discovered that employees feel pressured to respond to work-related emails outside of their designated working hours, adding further mental strain. However, effects have also been found to be more nuanced. Research by \cite{zahoorTechnologicalInnovationEmployee2022} indicates that employee well-being initially increases with technological innovation but reaches a threshold beyond which the positive effects diminish. Similarly, \cite{johnsonReviewAgendaExamining2020} suggests that while technology usage can increase the demands placed on employees and negatively impact their mental health, it can also have positive effects if it leads to more creative and meaningful work. Furthermore, \cite{zeikeDigitalLeadershipSkills2019} highlight the importance of managerial actions, as better digital leadership skills are associated with higher levels of employee well-being. Overall, the impact of digitization on employee mental strain and well-being is complex, influenced by various factors such as work demands, the nature of tasks, and the leadership practices applied within establishments.

\begin{figure}[ht!]
\caption{Effect on mental workload}
\includegraphics[width=10cm, height=10.26cm]{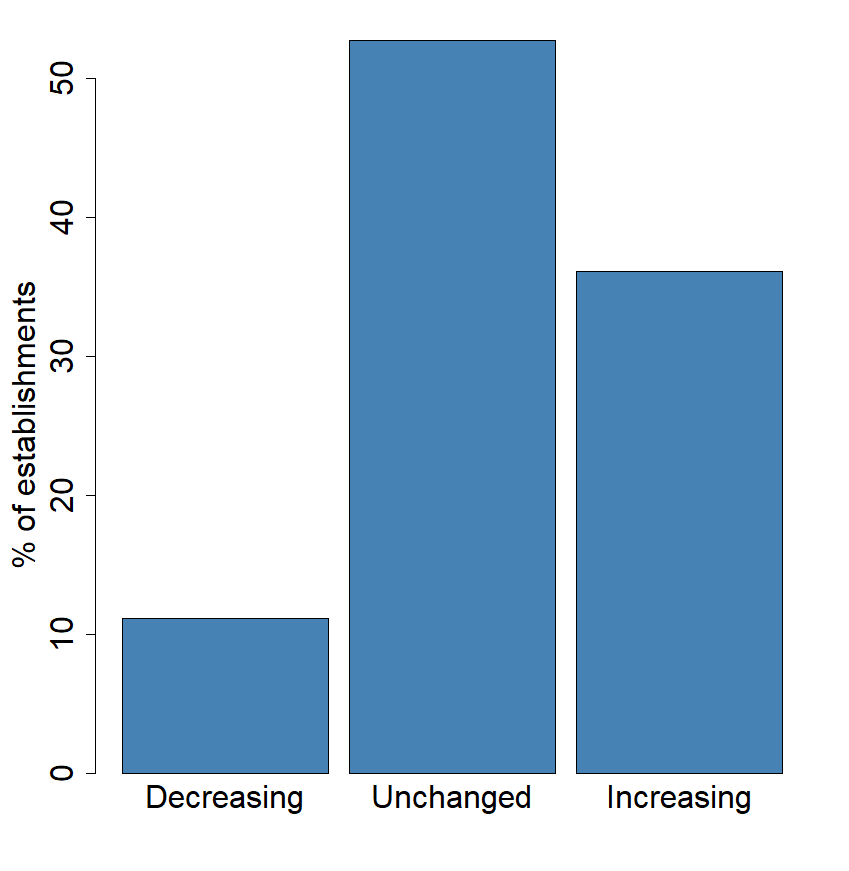}
\label{Psych}
\centering
\captionsetup{justification=centering,margin=1.5cm}
\caption*{\emph{Source: Swiss Employer Panel, 2022; own calculation and depiction.\\ Number of included observations: 404.}}
\end{figure}

Figure \ref{Psych} depicts the expected effect of the digital transformation on the mental workload of employees. While more than half of the establishments do not anticipate a change, 36\% expect an increase in the mental workload due to digitization. Only 11\% of the establishments expect a decrease in the mental workload. For establishments expecting an increase in the mental workload, it is important to consider the implementation of accompanying measures to prevent financial burden for businesses and costs for society associated with psychological distress, which can lead to health risks such as depression and burnout.

\begin{figure}[ht!]
\caption{Effect on mental workload: Divided by sector}
\includegraphics[width=10cm, height=10.26cm]{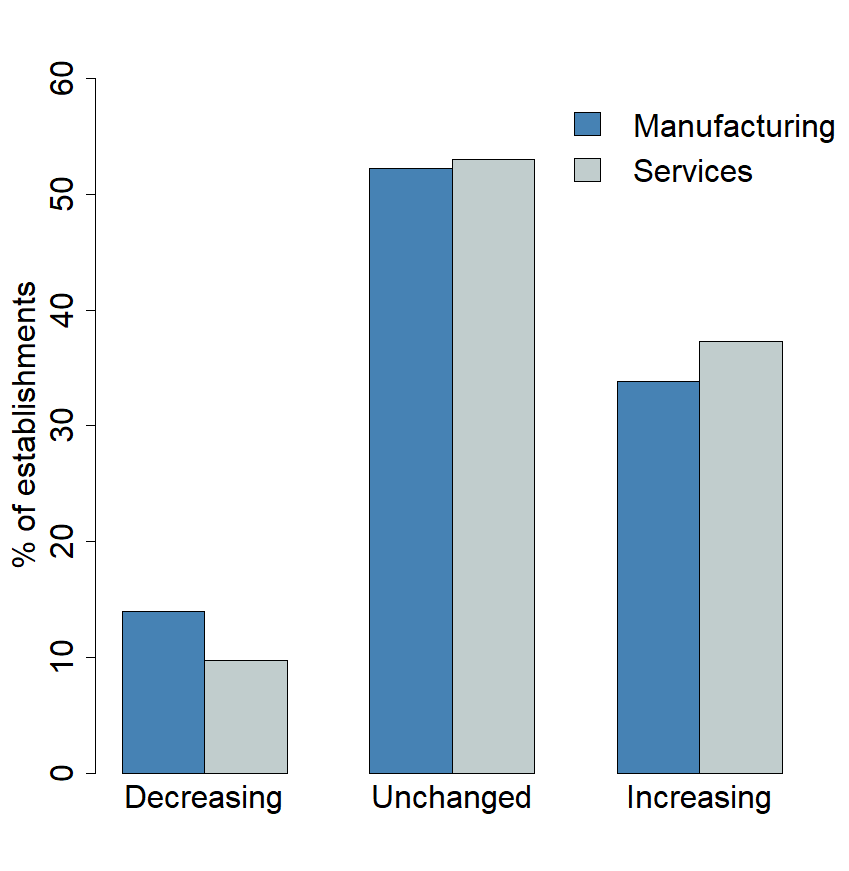}
\label{Psych_Sect}
\centering
\captionsetup{justification=centering,margin=1.5cm}
\caption*{\emph{Source: Swiss Employer Panel, 2022; own calculation and depiction.\\ Number of included observations: 404.}}
\end{figure}

Figure \ref{Psych_Sect} illustrates differences in expected changes concerning the mental workload by sector. Establishments in the manufacturing sector are more likely to anticipate a decrease in the mental workload, while establishments in the service sector are more likely to expect an increase. This difference can be interpreted in different ways. Firstly, it is possible that establishments in the service sector are more aware of potential mental health issues and therefore anticipate an increase in the mental workload. However, the finding can also be viewed in a more positive light, suggesting that the applied digital technologies in the manufacturing sector are more effective in reducing the mental workload, as they, for example, reduce stress associated with meeting deadlines or take over tedious and even dangerous manual labor.

There is no difference visible in the anticipated impact of digital technologies on the mental workload across establishments of different sizes. However, establishments that extensively employ digital technologies are more prone to anticipate an increase in the mental workload experienced by their employees than establishments which apply relatively little technologies.

\section{Conclusion and recommendations}
This paper provides a comprehensive descriptive overview of the current state of digital transformation in the Swiss economy. It adds to the existing literature by incorporating contemporary data and extensions of the research area.

Concerning the state of digitization, our investigation suggests that establishments could potentially benefit from implementing well-established technologies, as even fundamental and well-known technologies are not universally adopted. This approach can enhance an establishment’s
competitive position without incurring the high costs and uncertainties associated with cutting-edge technologies.

The implementation of new technologies rests on two pillars - the need for appropriate monetary funds to invest in a technology and the presence of the appropriate technological know-how in an establishment. The workforce gains technological know-how through appropriate further-training measures or strategic staffing decisions.

Further results indicate that many establishments are already aware of the importance of technological expertise. They expect an increase in the demand for further training measures as well as in the self-responsibility of employees to keep their knowledge up to date. With the growing demand for further training measures, taking a proactive role in addressing potential knowledge gaps within an organization will be critical.

Establishments expect not only an increase in the demand for employee knowledge but they also anticipate changes in the work flow. Work tasks are expected to become increasingly diverse and complex, while the number of simultaneous work tasks is also expected to rise. This development coincides with an increase in the mental workload for employees. Appropriate leadership practices are necessary to balance the increasing demands, such that employees are able to bring their best performance.

Furthermore, our analysis indicates that many establishments have an inflated perception about their technological state in comparison with their competitors. Recognizing a possible disparity can lead to a more realistic assessment of their competitive position and serve as a basis for making informed decisions regarding their strategic direction. Realizing that the competitive position is not as favorable as previously assessed is crucial to not miss out on potential benefits of the digital transformation.

Many establishments do not seem to be aware that the technological development influences both their internal and external transaction costs. Reevaluating core competencies and reassessing outsourcing and integration decisions can significantly enhance an establishment’s profitability.

\newpage

\bibliography{WP1}

\end{document}